\documentclass[sigconf,nonacm,9pt]{acmart}
\renewcommand\footnotetextcopyrightpermission[1]{}
\usepackage[utf8]{inputenc}
\usepackage{amsmath,amsfonts}
\usepackage{mathtools}

\usepackage{algorithm,algorithmic}
\usepackage{comment}

\usepackage[export]{adjustbox}
\usepackage{multirow}
\usepackage{xspace}

\newcommand{\sparqle}{\texttt{SPARQLe}\xspace}

\usepackage{caption}

\title{SPARQLe: \textbf{\underline{S}}ub-\textbf{\underline{P}}recision \textbf{\underline{A}}ctivation \textbf{\underline{R}}epresentation for \textbf{\underline{Q}}uantized \textbf{\underline{L}}LM Inferenc\textbf{\underline{e}}}

\author{
            Aradhana Mohan Parvathy$^1$, Soumendu Kumar Ghosh$^2$, Shamik Kundu$^2$, Arnab Raha$^2$, \\Souvik Kundu$^2$, Deepak A. Mathaikutty$^2$, Anand Raghunathan$^1$\\
$^1$\textit{Purdue University}, USA,
$^2$\textit{Intel Corporation}, USA
}

\thanks{$^{*}$Purdue University, West Lafayette, IN, USA (email: amohanpa@purdue.edu, raghunathan@purdue.edu).}%
\thanks{$^{\dagger}$Intel Corporation (email: soumendu.ghosh@intel.com,shamik.kundu@intel.com, arnab.raha@intel.com, souvikk.kundu@intel.com deepak.a.mathaikutty@intel.com). \\ This work was supported in part by CoCoSys, one of seven centers in JUMP 2.0, a Semiconductor Research
Corporation (SRC) Program sponsored by DARPA.
}%

\date{March 2022}
\usepackage{enumerate}
\usepackage{enumitem}

\setcopyright{none}

\begin{document}
\sloppy
\begin{abstract}

The rapid growth in sizes of Large language models (LLMs) results in high compute and memory costs during inference. Quantization has been a significant pathway to addressing this challenge. In the quest to push the limits of quantization, weights, which are static, can often be quantized aggressively ({\em e.g.} 4 bits), while activations often require higher precision ({\em e.g.,} 8 bits) to preserve accuracy, forcing hardware to operate with higher-precision datapaths. We leverage the statistical property that a significant fraction of activations are concentrated around zero, resulting in sparsity in the higher-order bits. Our proposal, SPARQLe, is a hardware–software co-design framework that exploits this sub-precision redundancy in any given quantized model. SPARQLe represents each 2k-bit activation tensor as a dense k-bit LSB tensor and a sparse k-bit MSB tensor compressed with a precision bitmap, and proposes a lightweight algorithm to increase MSB sparsity. SPARQLe reduces activation memory traffic and enables efficient computation on k-bit datapaths while preserving 2k-bit activation accuracy. SPARQLe includes an accelerator that operates directly on this hybrid format with minimal control overheads. Across the BitNet 3B, Llama2 7B, and Llama3 8B models, SPARQLe reduces prefill latency by \textbf{16-24.3\%} and decode latency by \textbf{13.5-23.4\%}, with \textbf{17-26.7\%} and \textbf{6.5-14.2\%} lower prefill and decode energy, respectively. SPARQLe demonstrates that sub-precision activation sparsity offers an effective and complementary pathway towards efficient LLM inference.
\end{abstract}

\maketitle 

\vspace{-0.1in}
\section{Introduction} \label{sec:Intro}

\begin{figure}[t]
    \centering
    \includegraphics[width=0.90\columnwidth]{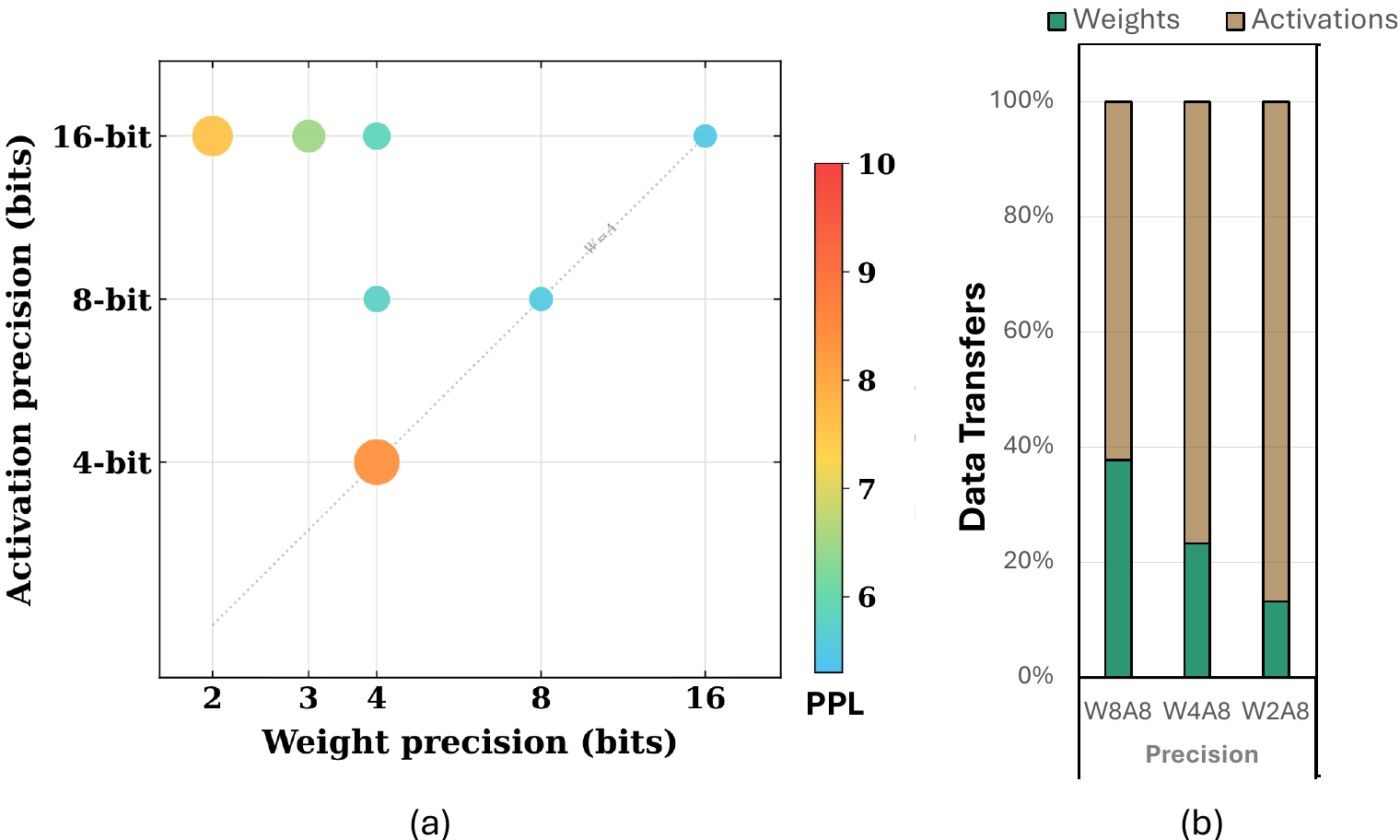}
    \caption{ (a) Perplexity for varying weight precisions (x-axis) and activation precisions (y-axis) for the Llama2-7B model \cite{omniquant,quarot,tesseraq,qserve,smoothquant}; bubble color indicates the perplexity increase relative to an FP16 baseline (lower is better). (b) proportion of data transfers due to weights and activations in Llama3-8B model across different precisions during prefill.}
    \label{fig:WAMotivation}
\end{figure}

Large language models (LLMs) have become central to modern AI systems, but their compute and memory demands continue to escalate, placing increasing pressure on the efficiency of hardware platforms. Quantization has emerged as a widely adopted approach to address this challenge. Initial quantization efforts successfully reduced both weights and activations from 16 bits to 8 bits with minimal accuracy loss~\cite{smoothquant}. However, pushing precision further reveals a fundamental asymmetry: while weights tolerate aggressive reduction to 4-bit precision or below~\cite{bitnet,qserve}, activations are far more sensitive— often requiring 8-bit precision to preserve model quality~\cite{qserve}. As shown in \figureautorefname~\ref{fig:WAMotivation}(a), reducing activation precision below 8 bits causes significant perplexity degradation for Llama3-8B even as weights may be further quantized. This asymmetry imposes significant efficiency costs on modern accelerators: on the compute side, processing 4-bit weights paired with 8-bit activations still utlizes 8-bit datapaths on current hardware, preventing weight quantization from translating into proportional reductions in compute energy and latency. On the other hand, activations constitute a predominant share of data movement as weight precision is minimized (\figureautorefname~\ref{fig:WAMotivation}(b)). Together, these bottlenecks make activation efficiency a critical challenge for scalable LLM inference. 

It is known that LLM activations typically follow Gaussian- or Laplacian-like distributions with a significant fraction of values concentrated near zero~\cite{activationDistribution}. In the typical case of 8-bit activations, this distribution causes a large fraction values to effectively lie within the 4-bit range, with their most significant 4 bits (MSB4) consisting of zeros or sign-extension bits. For instance, in Llama2-7B quantized to W4A8, over 58\% of activations entering the down-projection layer have MSB4 equal to zero. This sub-precision redundancy in quantized models has not been exploited for compute and data-movement efficiency. This is not sparsity in the conventional sense—individual activation values are nonzero—but rather \textit{sub-precision sparsity}: redundancy at the sub-operand level that existing dense or sparse accelerators cannot exploit~\cite{flexnn}, and that bit-serial~\cite{bitl} designs address only at the cost of substantial area and power overheads. Our proposal, \sparqle, takes a fundamentally orthogonal and complementary approach --- it exploits sub-precision sparsity within activations to achieve efficiency gains without altering the quantization scheme.

\sparqle is a hardware–software co-design framework that exploits sub-precision sparsity to enable efficient quantized LLM inference under asymmetric ({\em e.g.}, W4A8 or W2A8) representations. \sparqle decomposes each activation tensor into (i) a dense LSB4 tensor storing the least significant 4 bits of all elements, and (ii) a sparse MSB4 tensor compressed to retain only nonzero values via a precision bitmap (PBM). This decomposition creates distinct dense and sparse computations that cannot be efficiently exploited in tandem by existing accelerators, motivating a hybrid accelerator with a unified datapath supporting both modes. A lightweight algorithm further increases MSB4 sparsity through parameterized clipping, providing a fine-grained, controllable accuracy-efficiency tradeoff. 
To the best of our knowledge, \sparqle is the first work to exploit unstructured sub-precision activation sparsity, along with a hardware architecture that directly accelerates it. \sparqle is complementary to quantization methods and can benefit any quantized model with asymmetric weight {\em vs.} activation precision. In summary, we make the following contributions:
\begin{itemize}[leftmargin=*,itemsep=2pt,topsep=2pt]
\item \textbf{Sub-precision activation representation.} We propose a structured decomposition of 8-bit activations into dense LSB4 and sparse MSB4 components, along with a lightweight algorithm with controllable accuracy-efficiency tradeoffs, facilitating reductions in both compute and data-movement energy.
\item \textbf{Hardware–software co-design.} We develop a hybrid accelerator that performs dense and sparse computations directly on the proposed representation, eliminating costly decompress–compute–recompress cycles and delivering improvements in both latency and energy efficiency.
\item \textbf{Complementary to quantization.} \sparqle operates on already-quantized models without altering the underlying quantization scheme, making it directly composable with existing methods. Accuracy is governed by fine-grained clipping hyperparameters rather than fixed precision reduction, and efficiency gains stack on top of existing benefits from quantization.
\end{itemize}

\section{Background and Related Efforts} \label{sec:prelims}

This section provides a background on LLM inference  and reviews prior work on quantization.

LLM inference consists of two phases: prefill, which is compute-bound, and decode, which is memory-bound~\cite{kang2024gear}. The standard latency metrics for these phases are time-to-first-token (TTFT) and time-per-output-token (TPOT), respectively, which we also use to evaluate \sparqle.

A key technique for improving inference efficiency is quantization \cite{smoothquant,awq,bitnet,atom}, which reduces computation and memory overhead by lowering the precision of weights and activations, with integer formats commonly used in resource-constrained edge systems. A consistent observation across prior work is that activations are significantly more sensitive to quantization than weights \cite{bitnet,atom,awq}, requiring higher precision to maintain accuracy (\figureautorefname~\ref{fig:WAMotivation}(a)), leading to widely used configurations such as W4A8~\cite{bitnet,qserve}. 

Several works exploit the non-uniform distribution of tensor values to reduce the number of bits effectively needed for representation and computation. SqueezeLLM~\cite{squeezellm} uses dense-and-sparse decomposition to isolate outliers in \textit{static} weight tensors. However, activations are inherently \textit{dynamic}, varying across inputs and tokens, which makes static decomposition strategies ineffective and motivates techniques that exploit activation structure at runtime. On the hardware side, most programmable accelerators employ symmetric operand-width MAC units (e.g., $Int8\times
Int8$ or $Int4\times
Int4$) for flexibility to support diverse dataflows and workloads. When activations use higher precision than weights, computation is effectively dictated by the activation bit-width, limiting the benefits of weight quantization. While precision-reconfigurable~\cite{axbxp} and bit-serial~\cite{bitl} designs can address this mismatch, they introduce non-trivial overheads in control complexity, area, and power.

\sparqle takes a complementary approach. Instead of modifying the quantization method, it reduces the effective cost of high-precision activations by exploiting their statistical distribution. \textit{\sparqle is therefore complementary to existing quantization methods~\cite{blockdialect,microscopiq,bitnet,qserve, squeezellm}, operating on the representation of already-quantized activations.} While we focus on the common case of 8-bit integer activations for edge systems, the approach naturally extends to other precisions and floating-point quantized models.


\section{SPARQLe Methodology} \label{sec:methods}

This section presents \sparqle in three steps. We begin by outlining the activation data representation, followed by a description of the  method that increases sub-precision (MSB4) sparsity. We then introduce the \sparqle accelerator architecture, which leverages the hybrid data representation to achieve reductions in both latency and energy consumption.

\subsection{Activation Data Representation}
As discussed in Section \ref{sec:Intro}, the distribution of LLM activations causes many Int8 values to fall within the Int4 range, rendering their MSBs redundant. This can be systematically exploited to reduce latency and energy for both computation and data movement.
One possible approach is to adopt a mixed-precision strategy with two levels of precision: Int4 for small-magnitude values and Int8 for the remaining ones. However, storing activation tensors directly in this irregular mixed-precision form complicates memory accesses and degrades efficiency. To this end, \sparqle proposes a structured decomposition of activations that preserves efficiency while exploiting the inherent sparsity in MSBs.

\begin{figure}[t]
    \centering
    \includegraphics[width=0.98\columnwidth]{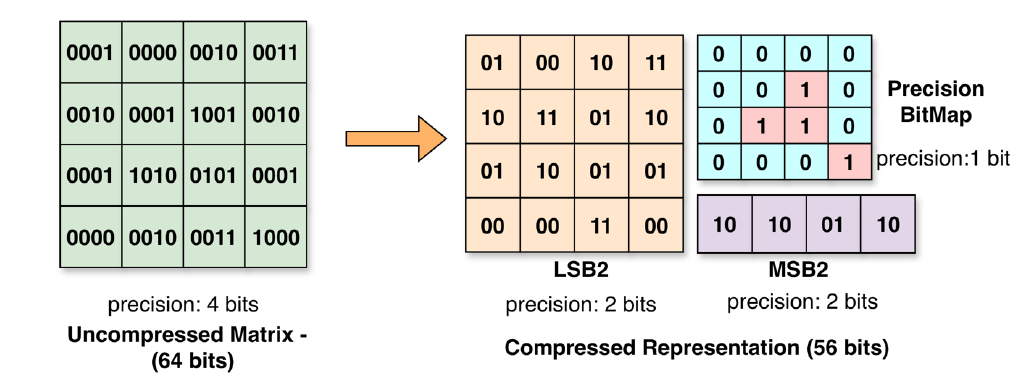}
    \caption{Illustration of the \sparqle data representation.}
    \label{fig:DataRep}
\end{figure}

Each activation tensor in \sparqle is decomposed into three structured components: (i) LSB4, a dense tensor storing the least significant 4 bits of all activation values; (ii) Precision BitMap (PBM), a binary mask, where 0 indicates that the corresponding value is fully represented by LSB4 and 1 indicates that the value requires MSB4; and (iii) MSB4, a sparse tensor storing the most significant 4 bits only for entries marked 1 in the PBM. \figureautorefname~\ref{fig:DataRep} illustrates this representation: for clarity, a 4-bit tensor is shown split at 2-bit granularity into LSB2, MSB2, and PBM; the same technique applies directly to 8-bit tensors decomposed into LSB4, MSB4, and PBM. Due to the distribution of LLM activations, MSB4 is naturally sparse. Compression is achieved by storing only nonzero MSB4 entries together with the PBM. For a p-bit tensor represented using this format, with s denoting the fraction of activations whose most significant p/2 bits are zero, the compression percentage is given by Equation~\ref{eq:compressionp}. In our work, we focus on p=8, corresponding to quantized 8-bit activations, in which case the compression percentage evaluates to $\frac{4s-1}{8}*100$. The proposed datapath (Subsection~\ref{subsec:Hardware}) operates directly on the LSB4, MSB4, and PBM representations without reconstructing full Int8 activations. 
This design skips unnecessary operations (\equationautorefname~\ref{eq:compute_red}) and data movement, lowering latency and energy consumption. 

\begin{align}
\scalebox{1.1}{$
\text{Compression (\%)} 
= \frac{p - (\frac{p}{2} + 1 + \frac{(1-s)p}{2})}{p} \cdot 100
= \frac{\frac{sp}{2}-1}{p} \cdot 100
$}
\label{eq:compressionp}
\end{align}

\vspace{-2ex} 

\begin{align}
\text{Ops Reduction (\%)} &= \frac{s}{2}*100
\label{eq:compute_red}
\end{align}

While most activations follow Gaussian- or Laplacian-like distributions, certain non-linear functions such as SiLU may cause activation distributions to deviate from these cases. Nevertheless, these activations still exhibit even higher sub-precision sparsity in practice — for instance, the SiLU output in Llama3-8B (decoder block 1) achieves 89\% sub-precision sparsity, compared to 32\% at the input to the q\_projection layer — making SPARQLe broadly applicable across activation types. For activations that are not naturally zero-centered, a zero-point adjustment during quantization can shift the distribution to increase sub-precision sparsity, making \sparqle applicable across a broad range of activation types. Further, as discussed next, this sparsity can be further enhanced through a lightweight algorithm without modifying any base model weights.


\subsection{Sub-Precision Sparsity Enhancement}\label{subsec:MSB4sparsityenhancement}

The effectiveness of \sparqle depends heavily on sub-precision sparsity, i.e., sparsity of MSB4. To further boost this sparsity, we introduce a lightweight algorithm that selectively clips activations into ranges where MSB4 is zero. The range of activation values for which MSB4 is equal to zero is defined by two bounds: the \textit{low-precision lower bound} (\(lp_l\)) and the \textit{low-precision upper bound} (\(lp_h\)). Activations in the range \([lp_l, lp_h]\) naturally have MSB4 equal to zero. 
For 8-bit integers in two’s complement representation, this range corresponds to \([0, 15]\). 
We extend this range by introducing the clipping constants \(l < lp_l\) and \(h > lp_h\). 
Specifically, values in \([l, lp_l)\) are clipped to \(lp_l\), and values in \((lp_h, h]\) are clipped to \(lp_h\). Values are clipped to these boundaries to keep them as close as possible to their original values, thereby minimizing error while increasing MSB4 sparsity. However, a naive application of this strategy to all activation tensors is impractical: exhaustive clipping introduces substantial error. Moreover, errors introduced in one layer do not remain localized; they propagate forward through the network, potentially compounding their effect. Hence, the clipping strategy must be applied selectively, carefully balancing the tradeoff between sparsity gains and accuracy loss.

\begin{figure}[t]
    \centering
    \includegraphics[width=0.85\columnwidth]{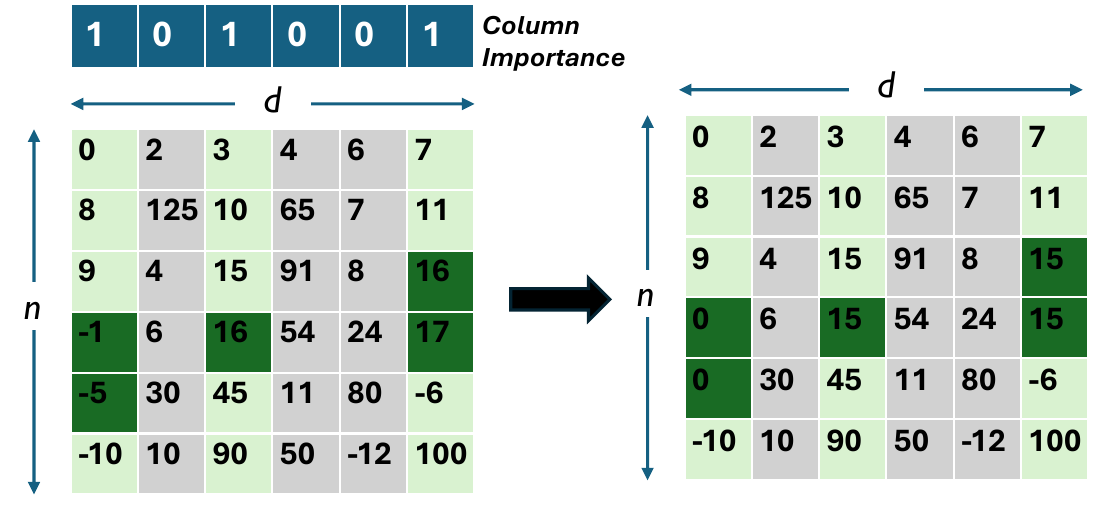} 
    \caption{ Example of MSB4 sparsity enhancement. Least-important columns (light green) are identified using precomputed weight-based column importance; values within the clipping range in these columns (dark green) are clipped.}
    \label{fig:Imposition}
\end{figure}

The impact of clipping-induced errors depends not only on the activation distribution but also on the magnitude of the weights with which they interact. Each value in a given activation column multiplies with the corresponding weight row values during matrix multiplication. Consequently, errors in a given activation column are scaled by the magnitudes of the weights in that row. To capture this effect, we define the \textit{importance} of a column as the $L_1$ norm of its associated weight row. Rather than applying clipping uniformly across the entire activation tensor, \sparqle imposes it selectively based on column importance. Specifically, only the least important $k$ columns-those whose associated weight rows have the smallest $L_1$ norms-are clipped. These low-importance columns are identified offline and stored as a binary mask, where a value of 1 indicates that the corresponding column is among the $k$ least important ones. This precomputation eliminates runtime overhead. By restricting clipping to low-importance columns, \sparqle introduces clipping where impact on accuracy is minimal, while still yielding substantial gains in MSB4 sparsity. \figureautorefname~\ref{fig:Imposition} illustrates an example of increasing MSB4 sparsity in an activation matrix using clipping constants $l=-5$ and $h=20$. The column-importance mask, precomputed from the weights of the corresponding layer, highlights the least important columns. In these columns, values falling within the clipping ranges are imposed on the boundary values $0$ and $15$, thus increasing the fraction of zero MSB4 entries while minimizing error.  

Having discussed the role of clipping in enhancing MSB4 sparsity, we now address the determination of the clipping constants $l$ and $h$. The choice of these constants critically influences the tradeoff between sparsity and accuracy, and we therefore consider two strategies for their selection: a global approach, which identifies a single set of constants for the entire model, and a fine-grained approach, which adapts the constants on a per-layer basis. In the global approach, a single set of clipping constants $[l,h]$ is identified and applied uniformly across all layers of the model. These constants are determined through calibration on a representative dataset: we sweep across candidate values of $l$ and $h$ and select the pair that yields the best calibration error/sub-precision sparsity tradeoff. However, this method cannot adapt to the heterogeneous sensitivity of different layers, motivating per-layer clipping constants.

Layer sensitivity depends on activation statistics and each layer’s impact on the final output. This variation motivates per-layer clipping constants instead of a single global choice to balance MSB4 sparsity and accuracy. For the fine-grained approach, we introduce clipping constants as trainable parameters within each linear layer. Let $\text{M}_{\text{base}}$ denote the original model without clipping, and $\text{M}_{\text{clip}}$ represent the model with learnable clipping constants. Let $\text{mask}_{L}$ denote the clipping mask (a binary mask indicating which activation values are clipped) for the activation matrix of layer $L$, determined by column importance and clipping constants. A higher number of ones in the mask corresponds to greater MSB4 sparsity improvement. The clipping constants are learned by training on a calibration dataset ($\mathcal{D}$) with all model parameters frozen except the constants (Algorithm \ref{algo:layerwise}), ensuring lightweight adaptation without altering the base model weights. The loss function used for training (line \ref{ref:loss}) combines two terms: (i) a mean square loss that measures the difference between the output of $\text{M}_{\text{base}}$ and $\text{M}_{\text{clip}}$, and (ii) a penalty term that encourages more aggressive clipping as shown in \equationautorefname~\ref{eq:loss}. The first term ensures that accuracy is preserved, while the penalty term encourages more MSB4 sparsity. This formulation enables the clipping constants to adapt automatically. In summary, the global method suits scenarios where no training is desired, while the layerwise method performs a minimal offline post-finetuning optimization—without modifying any base model parameters—to achieve higher sparsity and efficiency. 
\vspace{-0.1in}
\begin{equation}
\mathcal{L} = \text{MSELoss}\!\left(\text{M}_{\text{clip}}(\mathcal{D}),\,\text{M}_{\text{base}}(\mathcal{D})\right)
 - \alpha \sum_{L} \frac{1}{N_l}\sum_{i=1}^{N_l} \text{mask}_{L,i}
 \label{eq:loss}
\end{equation}


\subsection{SPARQLe Hardware Architecture}\label{subsec:Hardware}
\begin{figure*}[htbp]
    \centering
    \includegraphics[width=0.85\textwidth]{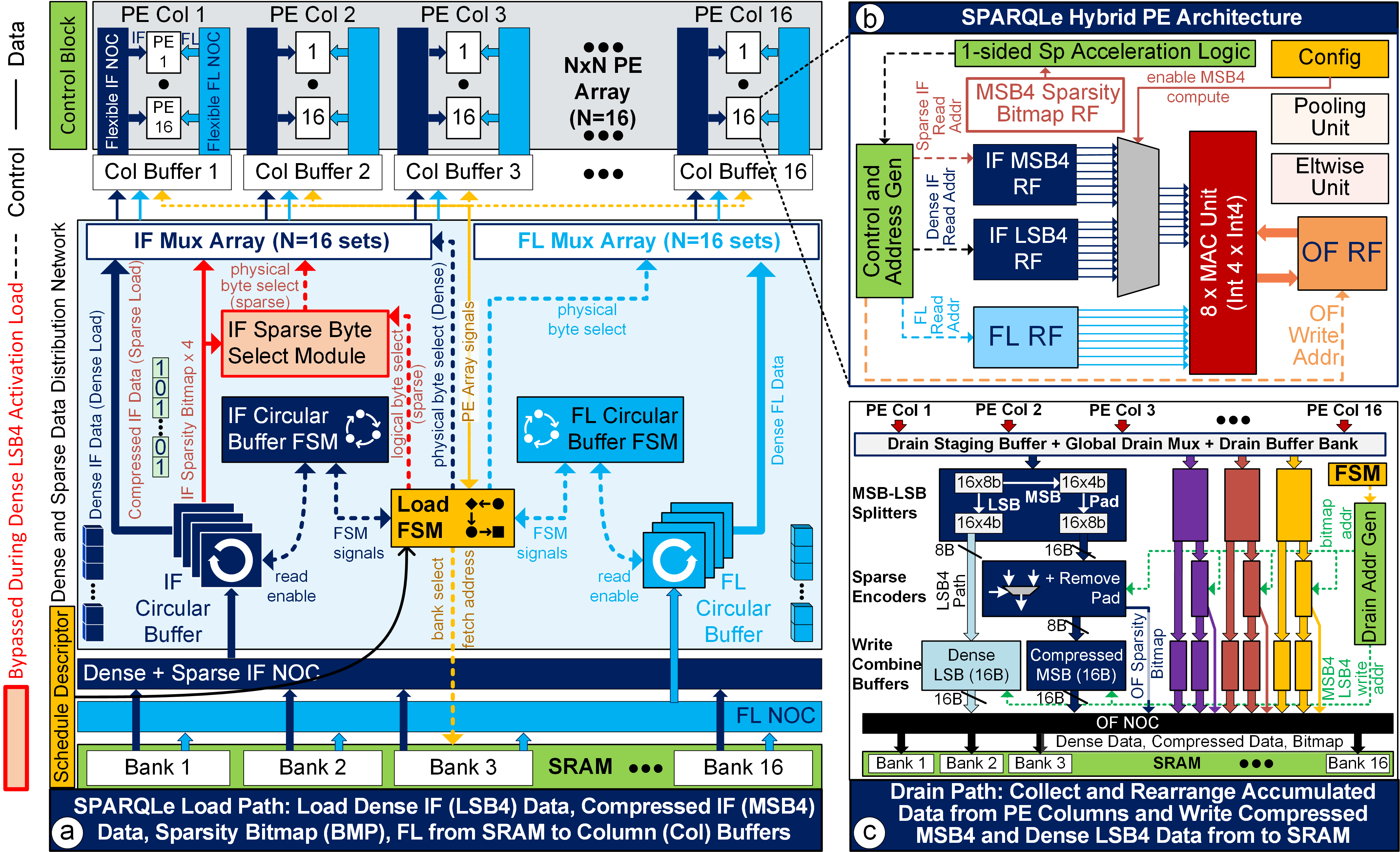}
    \caption{(a) Shared load path for dense and sparse phase, (b) Hybrid PE architecture, (c) Drain path in \sparqle.}
    \label{fig:load}
    \vspace{-15pt}
\end{figure*}
The decomposition of activations into dense \textbf{LSB4} and sparse \textbf{MSB4} components introduces computation patterns that existing dense or sparse accelerators fail to exploit efficiently. Conventional \textit{dense architectures} cannot leverage the sparsity-induced savings associated with MSB4 components, whereas \textit{sparse accelerators} incur substantial overhead—including metadata management, control complexity, and additional power consumption—when processing the dense LSB4 components. A design using separate dense and sparse datapaths duplicates buffers and MAC units while introducing nontrivial synchronization overhead, motivating sequential execution on a unified datapath.
Sequential execution of dense and sparse computations on a unified datapath fundamentally mitigates these limitations, as an output-stationary dataflow inherently supports accumulation of LSB4 and MSB4 partial sums without additional synchronization. Since \sparqle operates on LSB4 and MSB4 components at Int4 precision, it employs Int4 × Int4 MAC units rather than the Int8 × Int8 units required by a standard W4A8 dense accelerator. Under iso-area constraints, this allows \sparqle to instantiate $\sim$50\% more PEs than an Int8 dense baseline, partially offsetting the latency overhead of sequential execution. The remaining overhead is fully amortized once MSB4 sparsity exceeds $\sim$30\%; across the models evaluated in this work, the average MSB4 sparsity achieved ranges from 44\% to 62\%, comfortably exceeding this threshold across all layers and validating the sequential execution design choice.

To this end, we propose a hybrid accelerator that integrates dense and sparse datapaths within a unified processing element (PE) array. Each Int8 × Int4 or Int8 × Int2 operation is executed in two passes: (i) A dense pass operating on LSB4 activations; (ii) A sparse pass targeting MSB4 components, as indicated by the precision bitmap (PBM).
The hybrid PE array (\figureautorefname~\ref{fig:load}(b)) supports both computation modes without inducing irregular array-level data movement. The design draws inspiration from the two-sided sparse tensor-core–like architecture introduced in \cite{flexnn}, wherein sparsity logic locally identifies relevant inputs with minimal control and dataflow overhead. 
The subsequent subsections detail the load, compute, and drain phases of \sparqle’s execution pipeline.

    

{
\setlength{\textfloatsep}{1pt}
\setlength{\abovecaptionskip}{1pt}
\setlength{\belowcaptionskip}{2pt}
\begin{algorithm}[t]
\caption{Learning Layerwise Clipping Constants}
\label{alg:learn_clipping_constants}
\begin{algorithmic}[1]
\REQUIRE baseline model: $\text{M}_{\text{base}}$, learned model: $\text{M}_{\text{clip}}$, 
dataset: $\mathcal{D}$ 
\STATE Initialize clipping constants $\{\ell\}$ and $\{h\}$ for all linear layers
\STATE Freeze all weights in $\text{model}$ except $\{\ell\}$ and $\{h\}$
\FOR{each training epoch}
    \STATE $y_{\text{base}} = \text{M}_{\text{base}}(\mathcal{D})$; $y = \text{M}_{\text{clip}}(\mathcal{D},\ell,h)$
    \STATE Evaluate loss $\mathcal{L}$ \label{ref:loss}
    \STATE Backpropagate gradients of $\mathcal{L}$ and update $\{\ell\}$ and $\{h\}$ only
\ENDFOR
\STATE \textbf{return} $\{\ell\}$, $\{h\}$
\end{algorithmic}\label{algo:layerwise}
\end{algorithm}
}

\begin{figure}[t]
    \centering
    \includegraphics[width=\columnwidth]{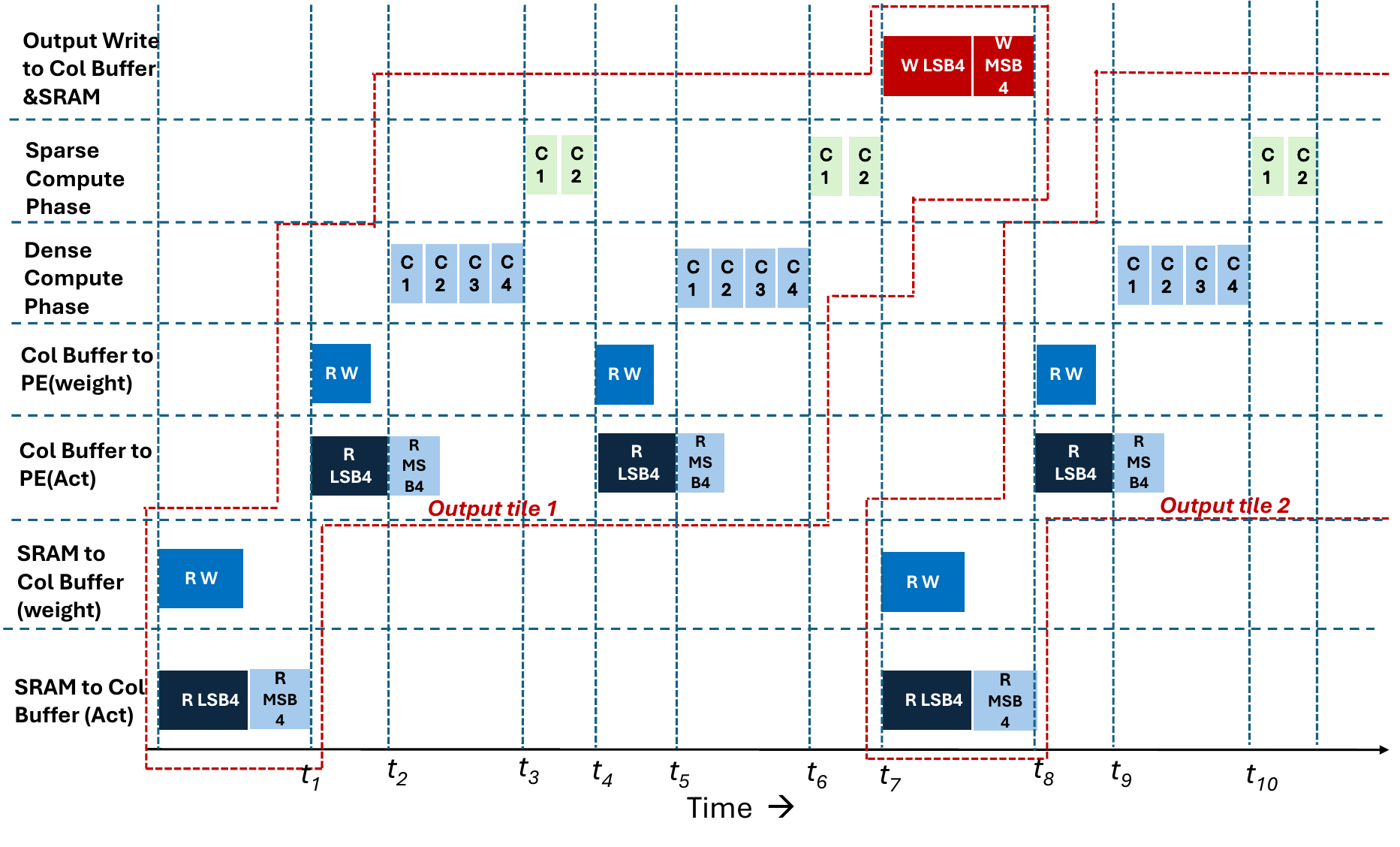} 
    \caption{Timeline of memory read/write, dense and sparse compute operations in \sparqle accelerator illustrating pipeline parallelism.}
    \label{fig:timeline}
\end{figure}

\textbf{Load Phase:} The load unit fetches weights and activations (LSB4, MSB4, and PBM) from SRAM and delivers them to the hybrid PE array through a hierarchical path: SRAM → circular buffers → column buffers → PE register files (RFs). A dedicated FSM generates addresses, bank-select signals, and coordinates the load across different buffers. The load datapath (\figureautorefname~\ref{fig:load}(a)) sustains a peak bandwidth of 32B/cycle from SRAM to circular buffers and 16B/cycle from column buffers to PE array.
These transfers occur independently for weights (broadcast across rows) and activations (broadcast across columns), while each PE receives 1 B/cycle for LSB4, MSB4, and weight operands.
Layer computation proceeds tile-by-tile to maximize reuse. To accommodate both dense LSB4 and sparse MSB4 activations, the load unit integrates a unified datapath with a bypass-able sparse-processing branch. During \textit{dense load phase}, LSB4 activations are read directly from the circular buffer. Subsequently, in \textit{sparse load phase}, a sparse-byte-select module facilitates loading of compressed MSB4 activations and PBMs.
This design enables the same load infrastructure to efficiently serve both dense and sparse activation flows, thereby avoiding hardware duplication. Compressed MSB4 values and PBMs are staged in column buffers and forwarded to RFs within each PE. Activations (IF) and weights (FL) are read concurrently (\figureautorefname~\ref{fig:timeline}, $t_0$-$t_1$) from SRAM/buffers, so the effective memory-access latency is determined by the slower of the two transfers. 

\textbf{Compute Phase:} \sparqle supports two execution modes: (i) dense LSB4 IF × dense FL and (ii) sparse MSB4 IF × dense FL. 
As shown in \figureautorefname~\ref{fig:load}(b), each PE integrates a shared Int4 $\times$ Int4 MAC unit, LSB4 RF, MSB4 RF, sparsity PBM RF, FL RF, and output RF, along with sparsity acceleration logic that uses PBM to select input operands (both FL and IF) corresponding to non-zero MSB4 IF \cite{flexnn}. Supported precisions include Int8×Int8, Int8×Int4, Int4×Int4, and Int4×Int2, requiring 4, 2, 1, and 1 compute rounds, respectively. During dense LSB4 compute, sparsity logic is bypassed; during sparse MSB4 compute, only valid operands are dispatched to the MACs. This enables a single PE to support both modes without hardware replication. The dense LSB4 compute phase executes first ($t_2 - t_3$ in \figureautorefname~\ref{fig:timeline}), followed by sparse MSB4 compute ($t_3 - t_4$  in \figureautorefname~\ref{fig:timeline}) using the shared MACs. Sparse partial sums are left-shifted by four bits and accumulated with the dense results inside the OFRF, thereby preserving  Int8 IF $\times$ FL numerical equivalence. Computation proceeds tile-by-tile, with column-buffer broadcasting data to the PE RFs initiating each tile's execution. Outputs are drained only after both dense and sparse computations (if present) for the current tile have been completed ($t_7 - t_8$ in \figureautorefname~\ref{fig:timeline}).
Although each compute stage internally consists of IF/FL RF reads, MAC execution, and OFRF writes—implemented in a pipelined manner—these microarchitectural details are abstracted in \figureautorefname~\ref{fig:timeline} for clarity. Note that MSB4 and PBM are written into IF RFs concurrently with the dense compute stage, avoiding any additional memory latency.


\textbf{Drain Phase:}
The drain unit transfers accumulated Int32 partial sums from the PE array to SRAM once both LSB4 and MSB4 tiled compute phases have completed. Outputs from the OFRF are drained through a hierarchy of local drain logic and special function units responsible for non-linear operations and Int8/Int4 quantization. These per-PE outputs then enter the global drain mechanism, which aggregates activations across all PE columns.
Our drain datapath (\figureautorefname~\ref{fig:load}(c)) adopts the global drain logic from~\cite{flexnn} up to the drain buffer bank (DB). Beyond DB, we introduce four MSB4–LSB4 splitters (one per four columns), which separate outputs into two parallel streams based on bit positions: (i) \textit{LSB4 path}: Lower 4 bits per activation do not require any processing and are written directly into the LSB4 write-combine buffer, providing 8B of LSB4 data per 16B of output per cycle. (ii) \textit{MSB4 path}: Upper 4 bits per activation are padded to 8-bits to enable sparse encoding. Sparse encoder~\cite{flexnn} generates both compressed MSB4 values and PBM. Compressed MSB4 data is written into the MSB4 write-combine buffer after the padding is removed. When both write-combine buffers are full, their contents are written to SRAM at an aggregate bandwidth of 32 B/cycle. LSB4 and MSB4+PBM streams are directed to separate SRAM banks to eliminate address conflicts.




\section{Experimental Methodology} \label{sec:exptMethod}

\textbf{Software}: We implemented \sparqle and evaluated it on integer-quantized LLMs. Specifically, we consider BitNet-3B (W2A8KV4) \cite{bitnet} and Llama2-7B and Llama3-8B \cite{llama2,llama3} quantized with \cite{qserve} (W4A8KV4). We use the  quantized models running on a traditional dense accelerator as the baseline for accuracy and efficiency. Experiments use the WikiText \cite{wikitext2} and LM-Harness \cite{lm-harness} datasets (with a maximum sequence length of 2048) for evaluation, with the WikiText \cite{wikitext2} dataset training split used for clipping-constant (low bound $l$, high bound $h$) calibration. 
For BitNet-3B, we calibrated layerwise clipping constants over 23 iterations on the calibration dataset, with all base model weights frozen. For Llama2 and Llama3, we adopt the global clipping constant calibration method, which integrates directly with the post-training quantization framework \cite{qserve} without requiring any changes. This demonstrates that \sparqle can be deployed with minimal global calibration effort and with layerwise calibration used depending on the available compute budget and user preference. We set $k$=50\% as a balanced operating point between no sparsity enhancement ($k$=0) and aggressive clipping across all columns ($k$=100\%), with the ablation in Section~\ref{sec:ablation} demonstrating that this choice yields a favorable accuracy--efficiency tradeoff. For accuracy comparison against W4A4 quantization, we use \cite{quarot} and \cite{duquant} as post-training quantization baselines for Llama2-7B and Llama3-8B respectively; BitNet-3B is excluded from this comparison as it is trained from scratch at low precision.



\noindent\textbf{Hardware}: We compared \sparqle with a baseline iso-MAC dense accelerator, following the design principles of \cite{flexnn} but without any sparsity-related modules across the architecture (Table~\ref{tab:hw_config}). Area and power estimates are derived from RTL synthesis using Synopsys Design Compiler targeting a 7nm process technology.
To evaluate performance, we developed an analytical energy–latency cost model similar to \cite{flexnn}. This model estimates per-layer energy and latency for:
(i) data movement across memory hierarchies, accounting for dense and sparse data with PBM overhead;
(ii) dense and sparse compute phases; and
(iii) end-to-end layer execution incorporating parallelism factors, including dense–sparse load, load–compute overlap (\figureautorefname~\ref{fig:timeline}), and pipelining.
For simplicity, multi-layer execution is modeled as sequential. Note that DRAM read/write energy and latency are excluded from the cost model generated results.

\begingroup
\setlength{\tabcolsep}{3pt}
\begin{table}[t]
\centering
\small
\caption{Baseline vs \sparqle accelerators hardware configuration}
\vspace{-0.05in}
\begin{tabular}{lcc}
\toprule
 & \textbf{Baseline} & \textbf{\sparqle} \\
\midrule
Memory Hierarchy      & 3 level   & 3 level \\
Number of Dense PEs   & 256       & 0 \\
Number of Hybrid PEs  & 0         & 256 \\
MAC precision         & Int4$\times$Int4 & Int4$\times$Int4 \\
MACs per cycle & 2048 & 2048 \\
RF (per PE)           & 224B      & 224B \\
On-chip buffer/SRAM   & 1.5MB     & 1.5MB \\
DRAM                  & 1GB       & 1GB \\
\bottomrule
\end{tabular}
\label{tab:hw_config}
\end{table}
\endgroup

\section{Experimental Results} \label{sec:Results}
This section illustrates the accuracy impact and performance improvement obtained using \sparqle. 
\begin{table}[htbp]
\centering
\vspace*{-0.1in}
\caption{Evaluation results across models and benchmarks. 
$\uparrow$ indicates higher is better, $\downarrow$ indicates lower is better.}
\resizebox{\columnwidth}{!}{
\begin{tabular}{l l c c c c c c c}
\toprule
\textbf{Model} & \textbf{Method} &
\textbf{PPL $\downarrow$} &
\textbf{BQ $\uparrow$} & 
\textbf{OQ $\uparrow$} & 
\textbf{PQ $\uparrow$} & 
\textbf{ARC-E $\uparrow$} &
\textbf{ARC-C $\uparrow$} &
\textbf{WGe $\uparrow$} \\
\midrule

\multirow{2}{*}\textbf{Llama3 8B}
& Baseline & 6.72 &80.55 & 35.2 & 78.34 & 80.72 & 52.05 & 73.00 \\
& \sparqle &7.05  &79.45 & 35.2 & 78.51 & 75.88 & 46.92 & 72.06 \\
\midrule

\multirow{2}{*}\textbf{Llama2 7B}
& Baseline & 5.67 &76.29 & 32.4 & 77.47 & 75.25 & 41.12 & 67.48 \\
& \sparqle & 6.0 &74.71 & 32.4 & 77.04 & 73.69 & 40.52 & 68.27 \\
\midrule

\multirow{2}{*}\textbf{BitNet 3B}
& Baseline & 10.21 &26.0 & 71.2 & 60.7 & 60.8 & 27.6 & 58.6 \\
& \sparqle & 12.19 &24.6 & 70.46 & 56.43 & 57.2 & 27.13 & 55.08 \\
\bottomrule
\end{tabular}
}
\label{tab:accuracy_results_together}
\vspace*{-0.1in}
\end{table}

\begin{figure*}[htbp]
    \centering    \vspace*{-0.1in}
\includegraphics[width=\textwidth]{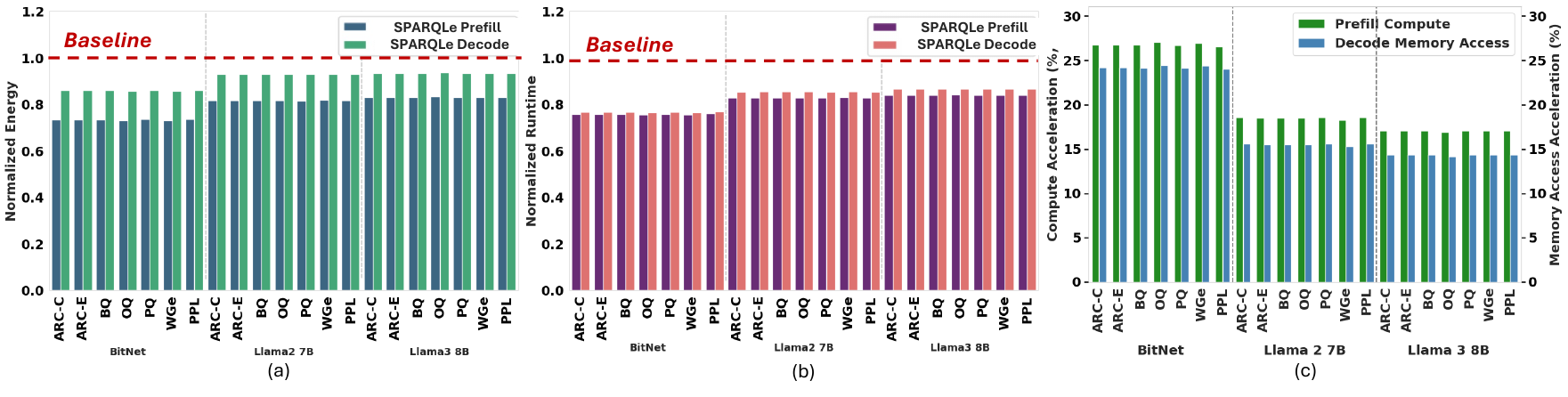} 
    \caption{Energy and performance benefits with \sparqle: (a) Normalized Prefill and Decode Energy, (b) Normalized Runtime Prefill and Decode, (c) Compute acceleration in Prefill and Memory access acceleration in Decode.}
    \vspace*{-0.15in}
    \label{fig:main_results}
\end{figure*}

\subsection{Accuracy and Performance Evaluation}
\begin{figure}[htbp]
    \centering    \includegraphics[width=0.9\columnwidth]{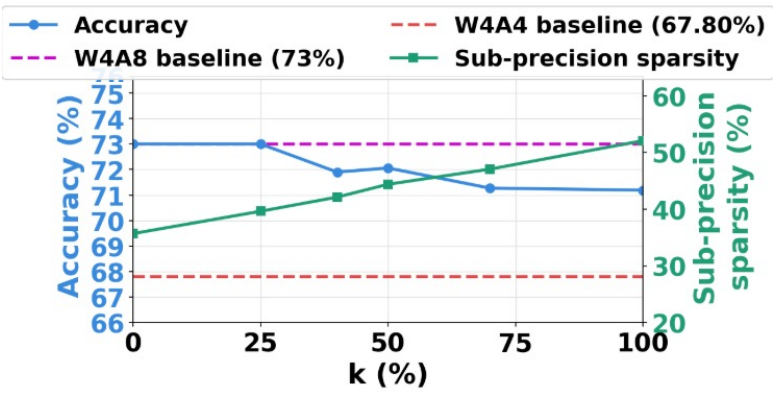} 
    \vspace{-0.1in}
    \caption{Accuracy/Sub-precision sparsity (averaged across the entire model) tradeoff across varying k in Llama3-8B model on WGe dataset}
    \label{fig:ablationk}
    \vspace{-1ex}
\includegraphics[width=0.9\columnwidth]{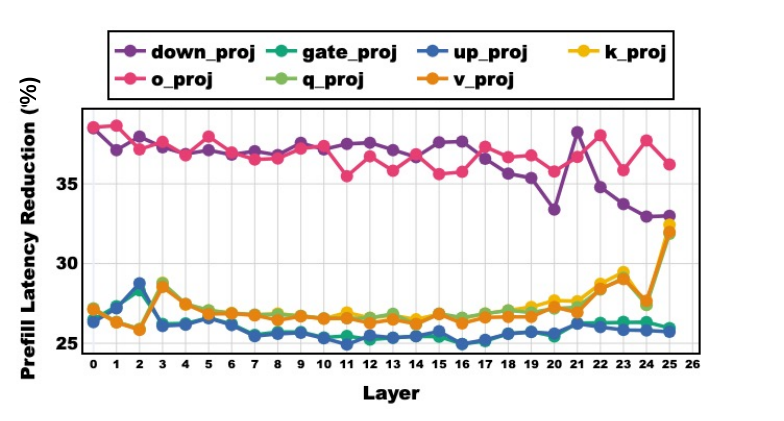} 
    \vspace{-0.1in}
    \caption{Layerwise latency reduction trend in BitNet-3B. Similar trend is observed for energy reduction as well.}
    \label{fig:ablation}
\end{figure}
Figures \ref{fig:main_results}(a) and \ref{fig:main_results}(b) summarize the end-to-end energy and performance benefits of \sparqle across the prefill and decode phases. \sparqle achieves average time-to-first-token (TTFT)/prefill speed-ups of \textbf{24.3\%}, \textbf{17.2\%}, \textbf{16\%}, and time-per-output-token (TPOT)/decode improvements of \textbf{23.4\%}, \textbf{14.6\%}, and \textbf{13.5\%} for BitNet-3B, Llama2-7B, and Llama3-8B respectively, with only minimal accuracy degradation across most of the benchmarks (Table \ref{tab:accuracy_results_together}). \textit{Despite small accuracy degradation in benchmarks such as ARC-C and ARC-E relative to the W4A8 baseline, \sparqle at W4A8 outperforms the corresponding W4A4 baselines on ARC-E by 3.82\% and 5.47\% for Llama2-7B and Llama3-8B respectively, and on ARC-C by 3.23\% for Llama3-8B, demonstrating that \sparqle occupies a favorable accuracy operating point between W4A8 and W4A4 while delivering efficiency gains without altering the underlying quantization scheme.} \sparqle also reduces inference energy in both phases, achieving average prefill energy savings of \textbf{26.7\%}, \textbf{18.4\%}, and \textbf{17.0\%} and decode energy savings of \textbf{14.2\%}, \textbf{7.1\%}, and \textbf{6.5\%} in BitNet-3B, Llama2-7B, and Llama3-8B, respectively. Across benchmark datasets, the resulting MSB4 sparsity averages 61.8\% in BitNet-3B, 47.0\% in Llama2-7B, and 44.4\% in Llama3-8B models. BitNet-3B exhibits the highest sparsity due to the lower weight precision (higher natural MSB4 sparsity) and the use of per-layer clipping constants, whereas the Llama models employ a global clipping constant that limits per-layer sparsity adaptation. 
Prefill benefits are consistently larger than decode benefits because SPARQLe reduces both data-transfer volume and compute cost for activations feeding linear layers, while activation–activation operations (e.g., $QK^{T}$, Softmax $\times$ V) are unaffected. Since prefill is both compute-bound and dominated by linear-layer activation movement, these gains translate more prominently into latency and energy improvements than in the memory-bound decode phase. Across all the models and benchmarks, \sparqle provides compute acceleration of \textbf{16.9\%} - \textbf{27.1\% }and data-transfer acceleration of \textbf{14.2\%} - \textbf{24.4\%} in the prefill and decode phases, respectively (\figureautorefname~\ref{fig:main_results}(c)). Unlike prior compression techniques that primarily target data-transfer reductions, \sparqle jointly reduces data-transfer cost (via compressed MSB4 representation) and compute cost (via sub-precision sparse execution). This combination produces consistent latency and energy gains in both prefill and decode phases across all evaluated models.

\subsection{SPARQLe Accelerator Area and Power}

For the configuration in Table \ref{tab:hw_config}, \sparqle incurs an area overhead of 5.5\% and an average power overhead of 7\% relative to the baseline. 
The area and power overhead relative to a 4-bit dense accelerator baseline arises from the additional sparsity logic added to support MSB4 sparsity. Despite the area and power overhead, \sparqle achieves substantial inference energy savings through runtime speedup and reduced data movement.

\subsection{Ablation Analysis}
\label{sec:ablation}

\noindent\textbf{Impact of $k$ on Accuracy--Sub-Precision Sparsity Tradeoff:} \figureautorefname~\ref{fig:ablationk} shows the effect of 
$k$ on sub-precision sparsity and accuracy, sweeping from $k=0$ (no sparsity enhancement) to $k=100$ (all activation columns amenable to clipping). At $k=0$, the network exhibits a natural sub-precision sparsity of 35.6\% averaged across layers, which can be further enhanced to 52\% with sub-precision sparsity enhancement. As $k$ increases, sub-precision sparsity exhibits an increasing trend at the cost of accuracy. Nevertheless, \sparqle consistently occupies accuracy operating points between W4A8 and W4A4 across the entire sweep, demonstrating that $k$ serves as a practical knob to navigate the accuracy-efficiency tradeoff without altering the underlying quantization scheme.

\noindent\textbf{Layerwise Latency and Energy Trends:} Since prefill exhibits higher latency and energy improvements than decode, we analyze the layerwise behavior of \sparqle in the prefill phase of BitNet 3B. \figureautorefname~\ref{fig:ablation} shows that benefits vary across layers, reflecting differences in MSB4 sparsity. Intermediate projections within the attention block and FFN — particularly \texttt{o\_proj} and \texttt{down\_proj} — exhibit higher gains, as their input activations follow more Laplacian-like distributions with sharper zero-centered peaks, yielding greater natural sub-precision sparsity compared to layers with Gaussian-distributed activations. This trend is consistent across all decoder blocks from early to deep layers. 


\section{Conclusion} \label{sec:Conclusion}
We present \sparqle, a hardware–software co-design framework for improving LLM inference efficiency under weight–activation precision asymmetry. \sparqle exploits the sub-precision structure of activations by decomposing them into dense LSB and sparse MSB components, enabling compressed representation and efficient execution. To further amplify these benefits, we introduce a lightweight sparsity-enhancement algorithm and a hybrid accelerator that operates directly on the decomposed format. Complementary to existing quantization methods, \sparqle delivers up to 25\% lower latency and 27\% higher energy efficiency.

\newpage
\balance
\bibliographystyle{ACM-Reference-Format}
\bibliography{ref}

\end{document}